\documentclass[longauth,traditabstract]{aa} 
\usepackage{amssymb}
\usepackage{graphicx}
\usepackage{color}
\usepackage{natbib}
\bibpunct{(}{)}{;}{a}{}{,} 
\newcommand{\nc}{\newcommand}
\nc{\lsun}{\ensuremath{\mathrm{L}_\odot}}
\nc{\msun}{\ensuremath{\mathrm{M}_\odot}}
\nc{\tex}{\ensuremath{\mathrm{T}_{\rm ex}}}
\nc{\cthree}{C$_3$}
\nc{\cthreehtwo}{$c$-C$_3$H$_2$}
\nc{\kms}{\mbox{km\,s$^{-1}$}}
\nc{\Kkms}{\mbox{K\,km\,s$^{-1}$}}
\nc\micron{\mbox{$\mu$m}}
\nc{\Trot}{$T_{\rm rot}$}%
\nc{\Ntot}{$N(C_3)$}%
\nc{\Tc}{$T_{\rm c}$}%
\nc{\Tdust}{$T_{\rm dust}$}%
\nc{\Tex}{$T_{\rm ex}$}%
\nc{\Tkin}{$T_{\rm kin}$}%
\nc{\Tmax}{$T_{\rm max}$}%
\nc{\cmcub}{\mbox{cm$^{-3}$}}
\nc{\cmsq}{\mbox{cm$^{-2}$}}
\newcommand{\HII}{H {\sc ii}}
\newcommand{\HI}{H {\sc i}}
\newcommand{\vlsr}{$v_{\rm LSR}$}
\newcommand\arcdeg{\mbox{$^\circ$}}%

\newcommand{\CCH}{\mbox{CCH}}             

\newcommand{\emm}[1]{\ensuremath{#1}}   
\newcommand{\emr}[1]{\emm{\mathrm{#1}}} 
\newcommand{\unit}[1]{\emm{\, \emr{#1}}}

\renewcommand{\deg}{\emm{^\circ{}}}
\newcommand{\Tsys}{\emm{T_\emr{sys}}}

\newcommand{\mm}  {\unit{mm}}

\newcommand{\MHz} {\unit{MHz}}
\newcommand{\GHz} {\unit{GHz}}

\begin{document}

\title{Detection of a dense clump in a filament interacting with W51e2
    \thanks{Based on data acquired with Herschel and IRAM observatories.
    Herschel is an ESA space observatory with science instruments provided 
    by European-led Principal Investigator consortia and with important 
    participation from NASA.}
}
   \author{
B.~Mookerjea\inst{\ref{tifr}}, 
C. Vastel\inst{\ref{irap1},\ref{irap2}},
G.~E.~Hassel\inst{\ref{siena}},
M. Gerin\inst{\ref{ens}}, 
J. Pety\inst{\ref{iram},\ref{ens}}, 
P.~F.~Goldsmith\inst{\ref{jpl}},
J.~H.~Black\inst{\ref{chalmers}}, 
T.~Giesen\inst{\ref{kosma}},
T.~Harrison\inst{\ref{siena}},
C.~M~.Persson\inst{\ref{chalmers}} 
J.~Stutzki\inst{\ref{kosma}} 
}

\institute{Tata Institute of Fundamental Research, Homi Bhabha Road,
Mumbai 400005, India \email{bhaswati@tifr.res.in}\label{tifr} 
\and
Université de Toulouse; UPS-OMP; IRAP; Toulouse, France\label{irap1}
\and
CNRS; IRAP; 9 Av. colonel Roche, BP 44346, F-31028 Toulouse cedex 4, France
\label{irap2}
\and 
Department of Physics and Astronomy, Siena College, Loudonville, NY
12211, USA\label{siena}
\and
LERMA, CNRS, Observatoire de Paris and ENS, France \label{ens}
\and
IRAM, 300 rue de la Piscine, 38406 St Martin d'Heres, France
\label{iram}
\and JPL, California Institute of Technology, Pasadena, USA\label{jpl}
\and  Onsala Space Observatory, Chalmers University of Technology, SE-43992 Onsala, Sweden \label{chalmers} 
\and
I. Physikalisches Institut, University of Cologne, Germany\label{kosma}
}

  \date{Received \ldots accepted \ldots}

\abstract
{In the framework of the Herschel/PRISMAS Guaranteed Time Key Program,
the line of sight to the distant ultracompact \HII\ region W51e2 has
been observed using several selected molecular species. Most of the
detected absorption features are not associated with the background
high-mass star-forming region and probe the diffuse matter along the
line of sight. We present here the detection of an additional narrow
absorption feature at $\sim$ 70 km~s$^{-1}$ in the observed spectra of
HDO, NH$_3$ and \cthree. The 70\,\kms\ feature is not uniquely
identifiable with the dynamic components (the main cloud and the
large-scale foreground filament) so-far identified toward this region.
The narrow absorption feature is similar to the one found toward
low-mass protostars, which is characteristic of the presence of a cold
external envelope.  The far-infrared spectroscopic data were combined
with existing ground-based observations of $^{12}$CO, $^{13}$CO, CCH,
CN, and C$_3$H$_2$ to characterize the 70\,\kms\ component. Using a
non-LTE analysis of multiple transitions of NH$_3$ and CN, we estimated
the density ($n(H_2)\sim$(1--5)\,10$^5$\,\cmcub) and temperature
(10--30\,K) for this narrow feature.  We used a gas-grain warm-up based
chemical model with physical parameters derived from the NH$_3$ data to
explain the observed abundances of the different chemical species. We
propose that the 70\,\kms\ narrow feature arises in a dense and cold
clump that probably is undergoing collapse to form a low-mass protostar,
formed on the trailing side of the high-velocity filament, which is
thought to be interacting with the W51 main cloud.  While the fortuitous
coincidence of the dense clump along the line of sight with the
continuum-bright W51e2 compact HII region has contributed to its
non-detection in the continuum images, this same attribute makes it an
appropriate source for absorption studies and in particular for ice
studies of star-forming regions.}


  \titlerunning{\cthree\ \& HDO in filament interacting with W51e2}
        \authorrunning{Mookerjea et al.}
   \maketitle

\section{Introduction}

The high-sensitivity large-scale CO maps of giant molecular clouds
provided the first evidence for the filamentary nature of the
interstellar medium \citep{ungerechts1987,bally1987}. The far-infrared
all-sky IRAS survey and several mid-infrared surveys (ISO, Spitzer/MIPS)
revealed the ubiquitous filamentary structure of both the dense and
diffuse ISM \citep{low1984}.  Most recently, the unprecedented
sensitivity and large-scale mapping capabilities of Herschel/SPIRE have
revealed a large network of parsec-scale filaments in Galactic molecular
clouds and have suggested an intimate connection between the filamentary
structure of the ISM and the formation of dense cloud cores
\citep{andre2010,molinari2010}. Here we present evidence of the
detection of a dense clump possibly formed by the interaction of an
extended foreground filament with the giant molecular cloud W51.

\begin{figure}
\begin{center}
\includegraphics[width=0.47\textwidth]{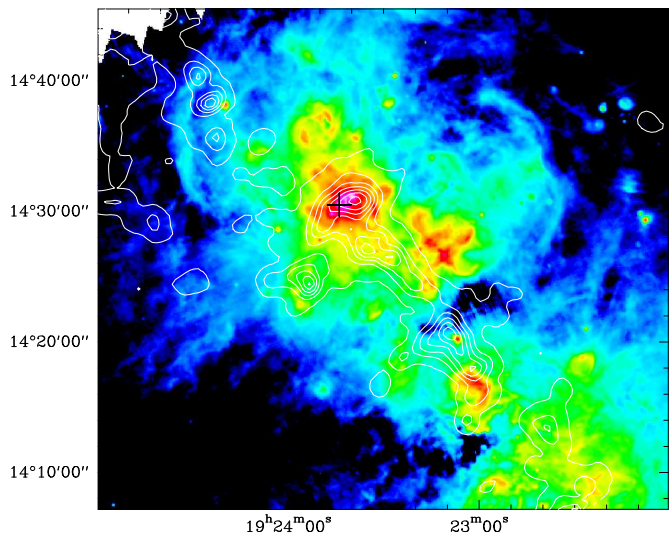}
\caption{Color image shows  70\,\micron\ continuum emission observed
with PACS. The white contours represent $^{13}$CO (1--0) emission
integrated between 68.9 and 70.3\,\kms\ corresponding to the
``filament", observed using FCRAO as part of the Galactic Ring Survey.
The black cross shows the position of the HIFI observations.
\label{fig_overview}}
\end{center}
\end{figure}

Located at a distance of 5.41$^{+0.31}_{-0.28}$\,kpc \citep{sato2010},
W51 is a radio source with a complicated morphology in which many
compact sources are superposed on extended diffuse emission
\citep{bieging1975}. The line of sight to W51 intersects the Sagittarius
spiral arm nearly tangentially (l = 49\arcdeg), which means that sources
over a  $\sim 5$\,kpc range of distances are superimposed on the line of
sight.  Based on the $^{12}$CO and $^{13}$CO 1--0 line emission of these
sources \citet{carpenter1998} divided the molecular gas associated with
the W51 HII region into two subgroups: a giant molecular cloud
(1.2$\times 10^6$ M$_{\odot}$) at $\sim$ 61~\kms, and an elongated
(22$\times$136\,pc) molecular cloud akin to a filamentary structure
(1.9$\times 10^5$ M$_{\odot}$) at 68~\kms.  While the brightest radio
source at 6\,cm (G49.5-0.4) is spatially and kinematically coincident
with the W51 giant molecular cloud, the G49.2-0.3, G49.1-0.4, G49.0-0.3,
and G48.9-0.3 radio sources seem to be associated with the 68~\kms\
cloud.  \citet{carpenter1998} speculated that the massive star formation
activity in this region resulted from a collision between the W51 giant
molecular cloud and the high-velocity (68~\kms) cloud.  G49.5-0.4 is the
brightest source in the W51 main region. The continuum emission from the
ultracompact \HII\ region was collectively named W51e before it was
resolved into compact components, labeled as W51e1 to W51e8
\citep{scott1978,gaume1993,mehringer1994,zhang1997}.
Figure~\ref{fig_overview} shows the 70\,\micron\ continuum emission as
observed with PACS by Hi-Gal in color, along with contours of $^{13}$CO
(1--0) emission integrated between 68.9 and 70.3\,\kms\ to locate the
GMC and the filament\footnote{The PACS data shown in
Fig.~\ref{fig_overview} are a Level 2 data product produced by the
standard data reduction pipeline for PACS-SPIRE parallel mode
observations as obtained from the Herschel Science Archive. The
$^{13}$CO data have taken from the Galactic Ring Survey done by FCRAO
available online at http://www.bu.edu/galacticring/}.  The continuum
emission appears to be better correlated with the filament, possibly
suggesting star formation activity in the filament.

As part of the {\em Herschel} key program ``PRobing InterStellar
Molecules with Absorption line Studies (PRISMAS) we have observed
several selected spectral lines in absorption toward the source W51e2 to
study the foreground material along the line of sight using HIFI.  These
observations have for the first time detected in absorption a component
at 70\,\kms\ (a velocity even higher than the velocity of the 68\,\kms\
filament) along with the 57\,\kms\ source velocity component, in
\cthree, HDO, and NH$_3$. The observation of \cthree\ was motivated by
the importance of small carbon chains for the chemistry of stellar and
interstellar environments either as building blocks for the formation of
long carbon chain molecules, or as are products of photo-fragmentation
of larger species such as PAHs \citep{radi1988,pety2005}.  \cthree\ has
been identified with {\em Herschel}/HIFI for the first time in the warm
envelopes of star-forming hot cores like W31C, W49N and DR21(OH)
\citep{mookerjea2010,mookerjea2012}.  Within the PRISMAS program, the
ground-state HDO transition at 894\,GHz has only been detected at the
velocity of the HII regions. Based on observations of high-mass
star-forming regions \citep{jacq1990,pardo2001, vandertak2006,
persson2007, bergin2010} and low-mass protostars \citep{fcliu2011,
coutens2012,coutens2013}, the HDO/H$_2$O ratio remains lower than the
D/H ratio derived for other deuterated molecules observed in the same
sources although it is clearly higher than the cosmic D/H abundance.
This low deuterium enrichment of water could provide an important
constraint to the astrochemical models to explain the chemical processes
involved in the formation of water.  Ammonia (NH$_3$) is a key species
in the nitrogen chemistry and  a valuable diagnostic because its complex
energy level structure covers a very broad range of critical densities
and temperatures. It has been observed throughout the interstellar
medium ever since its detection in 1968 \citep{cheung1968} but mainly in
its inversion lines at cm wavelengths.  With \emph{Herschel} it became
possible to observe ground-state rotational transitions with high
critical densities at THz frequencies of both ortho and para symmetries,
for instance, in cold envelopes of protostars \citep{hilyblant2010}, in
diffuse/translucent interstellar gas \citep{persson2012}, and hot cores
\citep{neill2013}.

We have combined the HIFI observations with IRAM observations of CCH,
\cthreehtwo\ and CN in all of which the 70\,\kms\ feature is detected in
absorption. We have used Local Thermodynamic Equilibrium (LTE) and
non-LTE models to derive the physical parameters of the dense clump at
70\,\kms\ and its relation to the W51 molecular cloud.  We also present
a gas-grain, warm-up chemical model that consistently explains the
abundances of all the species observed in the clump. This paper is
organized as follows: Sections 2 and 3 describe the Herschel and IRAM
observations, respectively. Section\,4 describes the complex velocity
structure observed along the sightline to W51e2. Section\,5 discusses
the column and volume densities derived from all the tracers that detect
the 70\,\kms\ clump. In Sec. 6 we compare the observed abundances of the
various chemical species with a grain warm-up based chemical model.
Section\,7 discusses various derived properties of the 70\,\kms\ clump
and presents a possible formation scenario for the clump.

\section{Observations with {\em Herschel}/HIFI}
\begin{table*}[h]
\begin{center}
\caption{Spectroscopic parameters for the observed spectral lines.\label{tab_allspec}}
\begin{tabular}{cccrrcccc}
\hline
\hline
Species & Telescope &Transition & Frequency & $E_{\rm low}$ & $A_{\rm ul}$   & Beam size  &  Reference \\
               &                     &                   &      GHz       &      K            &  s$^{-1}$  &                      &  \\  
\hline
\cthree, ($J$,$v$) & Herschel/HIFI & (9,1) -- (10,0) P(10)& 1654.081660 & 68.1 & 2.38~10$^{-3}$  & 12\arcsec\ & $a$ \\
                   & Herschel/HIFI & (3,1) --  (4,0) P(4) & 1787.890568 & 12.4  & 2.72~10$^{-3}$ & 12\arcsec\ & $a$ \\
                   & Herschel/HIFI & (2,1) --  (2,0) Q(2) & 1890.558057 & 3.7 & 7.51~10$^{-3}$   & 10\arcsec\ & $a$ \\
                   & Herschel/HIFI & (4,1) --  (4,0) Q(4) & 1896.706558 & 12.4  & 7.58~10$^{-3}$ & 10\arcsec\ & $a$  \\
c-C$_3$H$_2$, ($J_{K_a},{K_c}$) & IRAM-30m  &2$_{1,2}$--1$_{0,1}$   &
85.338893     & 2.35    &  2.32~10$^{-5}$      & 28\farcs3  & CDMS\\
CN          & IRAM-30m  &1,1/2,1/2--0,1/2,1/2   & 113.123370     & 0.0    &  1.29~10$^{-6}$      & 22\arcsec\   &  CDMS\\
            & IRAM-30m  &1,1/2,1/2--0,1/2,3/2   & 113.144157     & 0.0    &  1.05~10$^{-5}$      & 22\arcsec\   &   CDMS\\
            & IRAM-30m  &1,1/2,3/2--0,1/2,1/2   & 113.170492     & 0.0    &  5.15~10$^{-6}$      & 22\arcsec\   &   CDMS\\
            & IRAM-30m  &1,1/2,3/2--0,1/2,3/2   & 113.191279     & 0.0    &  6.68~10$^{-6}$      & 22\arcsec\   &   CDMS\\
            & IRAM-30m  &2,5/2,5/2--1,3/2,1/2   & 226.874191     & 5.5     &  9.62~10$^{-5}$      & 11\arcsec\   &   CDMS\\
            & IRAM-30m  &2,5/2,7/2--1,3/2,5/2   & 226.874781     & 5.5     &  1.14~10$^{-4}$      & 11\arcsec\   &    CDMS\\
            & IRAM-30m  &2,5/2,3/2--1,3/2,1/2   & 226.875896     & 5.5     &  8.59~10$^{-5}$      & 11\arcsec\   &    CDMS\\
            & IRAM-30m  &2,5/2,3/2--1,3/2,3/2   & 226.887420     & 5.5     &  2.73~10$^{-5}$      & 11\arcsec\   &    CDMS\\
            & IRAM-30m  &2,5/2,5/2--1,3/2,5/2   & 226.892128     & 5.5     &  1.81~10$^{-5}$      & 11\arcsec\   &    CDMS\\
            & IRAM-30m  &2,5/2,3/2--1,3/2,5/2   & 226.905357     & 5.5     &  1.13~10$^{-6}$      & 11\arcsec\   &    CDMS\\
CCH, N$_{J,F}$  & IRAM-PdBI  &1$_{3/2,1}$--0$_{1/2,0}$ &  87.328585  & 0.0  & 1.27~10$^{-6}$       & 5\farcs6 $\times$ 4\farcs6   &    CDMS\\
para-H$_2^{18}$O  & Herschel/HIFI   & 1$_{11}$-0$_{00}$  & 1101.698257 &
0.0   & 1.79~10$^{-2}$  & 19\farcs2  &JPL \\
ortho-NH$_3$      & Herschel/HIFI   & 1$_0^-$--0$_0^+$       &  572.498160   &  0.0     & 1.58~10$^{-3}$  & 37\farcs5  & JPL\\
                             & Herschel/HIFI   & 2$_0^+$--1$_0^-$       & 1214.852942  & 27.5    & 1.81~10$^{-2}$  & 17\farcs5  & JPL\\
para-NH$_3$       & Herschel/HIFI   & 2$^-_1$--1$^+_1$       & 1168.452394  & 23.3    & 1.21~10$^{-2}$  & 17\farcs5  & JPL\\
                             & Herschel/HIFI   & 2$^+_1$--1$^-_1$       & 1215.245714  & 22.2    & 1.36~10$^{-2}$  & 17\farcs5  & JPL\\
                             & Herschel/HIFI   & 3$_2^+$--2$_2^-$       & 1763.823186  & 64.5    & 3.30~10$^{-2}$  & 12\farcs  & JPL\\
HDO              & Herschel/HIFI   & 1$_{1,1}$--0$_{0,0}$ & 893.638666 &
0.0    & 8.35~10$^{-6}$ & 24\farcs1 & JPL\\
\hline
$a$ \citet{mookerjea2010}
\end{tabular}
\end{center}
\end{table*}

In the framework of the Herschel Key Program PRISMAS (P.I. M.  Gerin),
we observed the line of sight toward the compact HII region W51e2, whose
coordinates are $\alpha(J2000)$ =  19$^h$23$^m$43\fs90 $\delta(J2000)$=
14\arcdeg30\arcmin30\farcs5.  The observations were performed in the
pointed dual-beam switch (DBS) mode using the HIFI instrument
\citep{deGraauw2010,roelfsema2012} onboard {\it Herschel}
\citep{pilbratt2010}. The DBS reference positions were situated
approximately 3$\arcmin$ east and west of the source. The HIFI Wide Band
Spectrometer (WBS) was used with optimization of the continuum,
providing a spectral resolution of 1.1\, MHz  over an instantaneous
bandwidth of 4~$\times$~1\,GHz. To separate the lines of interest from
the lines in the other sideband, which could possibly contaminate our
observations, we observed the same transition with three different LO
settings. This method is necessary in such chemically rich regions to
ensure genuine identification of spectral lines.  The data were
processed using the standard HIFI pipeline up to level 2 with the
ESA-supported package HIPE 8.0 \citep{ott2010} and were then exported as
FITS files into CLASS/GILDAS format\footnote{See
\texttt{http://www.iram.fr/IRAMFR/GILDAS} for more information about the
GILDAS software.} for subsequent data reduction. Both polarizations were
averaged to reduce the noise in the final spectrum. The baselines
obtained and subtracted are well fitted by straight lines over the
frequency range of the whole band. The single sideband continuum
temperature derived from the polynomial fit at the line frequency was
added back to the spectra. 

We followed up the detection of the 70\,\kms\ velocity component in
\cthree\ by searching for the feature in different available transitions
of HDO, NH$_3$, CCH, \cthreehtwo, and CN.  Among these chosen molecules,
\cthreehtwo\ and CCH are chemically related to \cthree\ since they are
the by-products of the proposed formation route of \cthree\ via warm-up
of the CH$_4$ mantle of the grains, which essentially refers to a
temperature of 30\,K and higher \citep{hassel2011}. In contrast, NH$_3$
and HDO are tracers of cold (20\,K) and dense molecular gas.  The
abundance of CN is typically found to be low in hot cores, and higher in
regions of intense UV fields such as the interfaces of \HII\ regions and
molecular clouds.

\subsection{Linear carbon chain, \cthree}

We observed four transitions of the $\nu_2$ fundamental band,
$P(4)$, $Q(2)$, $Q(4)$, and $P(10)$ of triatomic carbon using the bands
7a, 7b, and 6b of the HIFI receiver. All spectroscopic information
regarding the observed \cthree\ lines is presented in
Table~\ref{tab_allspec}.  The observations were carried out on October
29, 2010 and April 22 and 23, 2011. Further details of the observations
and data reduction are similar to the analysis of the W31C and W49N
sources presented in \citet{mookerjea2010}.  All spectra  were smoothed
to resolutions of between $\sim$ 0.16 to 0.18~\kms\ and the rms noise level
for the spectra ranges between 0.03 to 0.04~K (Fig.\ref{fig_c3spec}).

\subsection{NH$_3$} 

We observed the ortho--NH$_3$ 2$_0^+$--1$_0^-$ and 1$_0^-$--0$_0^+$
ground-state transitions at 1214.8 (band 5a) and 572.5 GHz (band 1b) as
well as the para--NH$_3$ 2$_1^+$--1$_1^-$ ground transition at 1215.2
GHz (band 5a) \citep{yu2010}. The beam size is $\sim$37$^{\prime\prime}$
for the low frequency transition and 17.5$^{\prime\prime}$ for the
higher frequencies transitions. The forward efficiency is 0.96 and the
main beam efficiencies are 0.76 (at 572 GHz) and 0.64 (at 1215 GHz). The
PRISMAS observations were carried out on October 29, 2010 and April 22,
2011.  Additionally, we used the para--NH$_3$ 2$_1^-$--1$_1^+$
transition at 1168.45\,GHz and 3$_2^+$--2$_2^-$ transition at
1763.82\,GHz observations obtained within the OT1 programme on
absorption studies of N-hydrides. These observations were performed on
April 18 and 21, 2012.  Spectroscopic and observational details for all
transitions are summarized in Table~\ref{tab_allspec}.

\subsection{HDO}

We observed one of the HDO fundamental transitions
(1$_{1,1}$--0$_{0,0}$) at 893.6 GHz (using the spectroscopic catalog
JPL: \citet{pickett1998}) in band 3b. The beam is about 24\farcs1 and
the forward efficiency and the main beam efficiency are 0.96 and 0.74,
respectively.  The observations were carried out on October 27, 2010.
Spectroscopic and observational details are summarized in
Table~\ref{tab_allspec}.

\section{Ground-based IRAM observations}

\subsection{CCH with PdBI}

Plateau de Bure Interferometer (PdBI) observations dedicated to the
PRISMAS project were carried out with six antennas in September 2007.
The 4\GHz{} instantaneous IF--bandwidth allowed us to simultaneously
observe \CCH{} (plus six other lines) at 3\mm{} using seven different
40\MHz{}--wide correlator windows. One additional 320\MHz{}--wide
correlator window was used to simultaneously measure the 3\mm{}
continuum.  The CCH line was detected in emission and absorption
(spectroscopic details in Table\,\ref{tab_allspec}).

We observed W51e2 in track-sharing mode in the D configuration (baseline
lengths from 24 to 97~m). A two-fields Nyquist-sampled mosaic was needed
to cover the W51e2 continuum emission.  The D configuration was observed
for about 7~hours of \emph{telescope} time. This leads to
\emph{on--source} integration time of useful data of 1.5~hours for W51e2
after removing the bad data. The rms amplitude noise of the bandpass
calibration was 0.5\%.  The rms phase noise were between 7 and
18\deg{}, which introduced position errors of $<$0\farcs25.  The typical
resolution is 5\farcs1 and the primary beam at the frequency of the
CCH transition at 87.328585\,GHz is 56\farcs6. Table~\ref{tab_cchpdbi}
presents the relevant parameters for these observations.

The data were reduced with the GILDAS software supported at
IRAM~\citep{pety05b}.  Standard calibration methods using nearby
calibrators were applied. The calibrators used for the absolute flux
calibration are  B1749$+$096 (5.2 Jy) and  B1923$+$210 (1.4\,Jy),
respectively.  The W51e2 data were processed using natural weighting to
obtain the optimal signal-to-noise ratio. The line emission extension
exceeds one third of the primary beam, implying the need for
short-spacings from a single-dish telescope.  However, we are interested
here in the absorption lines in front of the continuum source, which is
compact enough to avoid the need for short-spacings. The data were
deconvolved using the H\"ogbom CLEAN algorithm with a support defined
around the continuum source to guide the search of the CLEAN components.

\begin{table*}
    \caption{Observation parameters for CCH with PdBI}
    \begin{center}
      \begin{tabular}{lrrccc}
        \hline
        \hline
        & \multicolumn{2}{c}{Phase center} & Number of fields & Field-of-view & Obs. date \\
        W51  &  $\alpha_{2000} = 19^h23^m43.40^s $ & $ \delta_{2000} = 14^\circ 30' 34.8''$ & 2 & $72''\times 58''$ & 7-sep-2007 \\
        \hline
      \end{tabular}
      \medskip{}
      \begin{tabular}{lcccccccc}
        \hline
        \hline
        & configuration & Beam   & PA     & Resol  & Water & \Tsys{} & On-source & Noise$^{a}$ \\
        &               & arcsec & \deg{} & \kms{} & mm    & K       &    hours & mJy/Beam\\
        W51  & D  & $ 5.61 \times 4.63 $ & 132.6 & 0.27 & 4 & 100 & 1.5 & 9\\
        \hline
      \end{tabular}\\
      $^{a}$ The noise values quoted here are those at the mosaic center
      (mosaic noise is inhomogeneous because of primary beam correction; it 
      steeply increases at the mosaic edges).\\
    \end{center}
    \label{tab_cchpdbi}
\end{table*}

\subsection{CN and \cthreehtwo\ with IRAM 30m}

The CN $J~=~1$--0 transition was observed at the IRAM-30 m telescope at
Pico Veleta (Spain) in December 2006 in wobbler-switching mode and
detected through its F~=~1/2--1/2, 1/2--3/2, 3/2--1/2, 3/2--3/2
hyperfine components (see Table \ref{tab_allspec}). All the details
about these observations can be found in \citet{godard2010}. The CN
$J~=~$2--1 transition was also observed in parallel with $J~=~$1--0 in
its F~=~5/2--3/2, 7/2--5/2, 3/2--1/2, 3/2--3/2, 5/2--5/2 and 3/2--5/2
hyperfine components (see Table \ref{tab_allspec}), although this is not
reported in \citet{godard2010}.  The half-power beam width is about
22$^{\prime\prime}$ at 113 GHz compared with the 11$^{\prime\prime}$
value at 227 GHz.

The c--C$_3$H$_2$ 2$_{1,2}$--1$_{0,1}$ transition at 85.3 GHz was
observed in December 2006 \citep{gerin2011}.  The spectroscopic details
are quoted in Table\,\ref{tab_allspec}.


\section{New velocity component along the W51e2 sightline }

To determine the spatial distribution of gas in the W51e2 region, the
velocity components originating in the local gas need to be separated
from those produced by gas along the line of sight to the background
continuum source.  Fortunately, this source was observed in detail using
several transitions of atomic and molecular species. For example,
H109$\alpha$ observations toward W51e2 showed that the average velocity
of the ionized gas is $\simeq 56.8$~\kms \citep{wilson1970}.
\citet{koo1997} used \HI\ observations toward W51e2 to identify two sets
of absorption features, the local features at 6.2, 11.8 and 23.1~\kms\
and features proposed to be belonging to the Sagittarius spiral arm at
51.1, 62.3, and 68.8~\kms.  We note that for the \HI\ observations the
line widths range between 7.1 and 12.2~\kms\  with a spectral resolution
of 2.1~\kms, thus leading to an uncertainty in the velocity
determination. However, the highest velocity permitted by Galactic
rotation is about 60~\kms\ in this direction. Therefore the 62.3 and
68.8~\kms\ components are most likely associated with a high-velocity
(HV) stream. Note also that the CO line emission is brightest at \vlsr\
= 58~\kms\ \citep{kang2010}, indicating that the ionized and molecular
gas have similar velocities and are probably associated with the same
radio continuum source. 

Using H$_2$CO observations, \citet{arnal1985} detected five absorption
components at $\sim$ 52, 55, 57, 66, and 70~\kms\ toward W51e2.  As
discussed above, the 57~\kms\ component seems to be associated with the
HII region.  The 62~\kms\ component detected in \HI\ and not in H$_2$CO
is understood to be a \HI\ cloud associated with W51e2 (G49.5-0.4).
H$_2$CO absorption between 66 and 70~\kms\  and HI absorption at
68.8~\kms\ imply that the HV stream is in front of W51e.  In the
large-scale $^{12}$CO and $^{13}$CO channel maps, the emission feature
in the 66 to 70~\kms\ velocity range traces a filamentary structure that
extends across a region $\sim$ 1$^{\circ}$ in size that contains most of
the \HII\ regions associated with W51.  This is consistent with the
triggered star formation scenario that is due to the interaction of the
filament with the main cloud, as proposed by \citet{carpenter1998}.
Recently, \citet{kang2010} compared the molecular cloud morphology with
the distribution of IR and radio continuum sources and found
associations between molecular clouds and young stellar objects listed
in Spitzer IR catalogs. Star formation in the W51 complex appears to
have occurred in a long and large filament along with the \HII\ regions.
Young stellar objects along the filament are younger and more massive
than those in the surrounding regions \citep{carpenter1998}.
Cloud-cloud collisions may compress the interface region between two
colliding clouds and initiate star formation. The molecular clouds at
this interface are heated, but the molecular clouds on the trailing side
remain cold. The diffuse emission from the giant molecular cloud
truncates at the location of the 68\,\kms\ cloud for velocities higher
than 63\,\kms. The W51 cloud and filament are probably at a common
distance since such an interface is unlikely to result from the chance
superposition of unrelated molecular clouds.


\begin{figure}
\begin{center}
\includegraphics[width=0.4\textwidth]{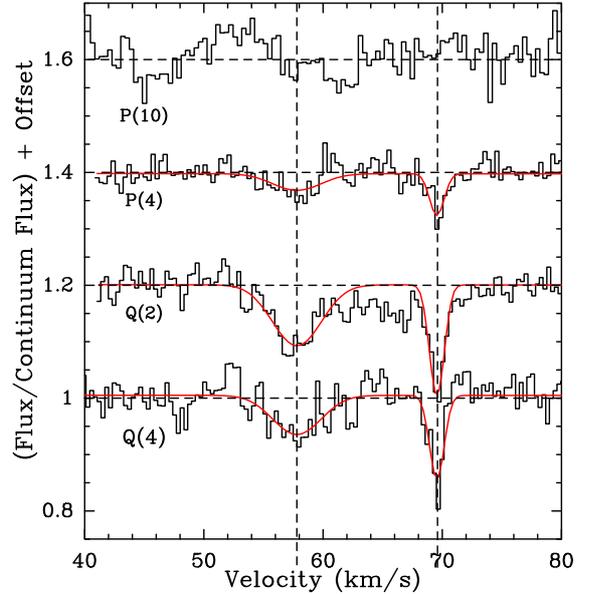}
\caption{ HIFI observations of the \cthree\ lines.
The observed spectra (in black) are corrected for the DSB continuum and
normalized to the SSB continuum level. Results of simultaneous fitting
of $P(4)$, $Q(2)$ and $Q(4)$ are shown in red.  The observed spectrum of
$P(10)$ is also shown to illustrate the non-detection.
\label{fig_c3spec}}
\end{center}
\end{figure}

\begin{figure*}
\begin{center}
\includegraphics[width=0.45\textwidth]{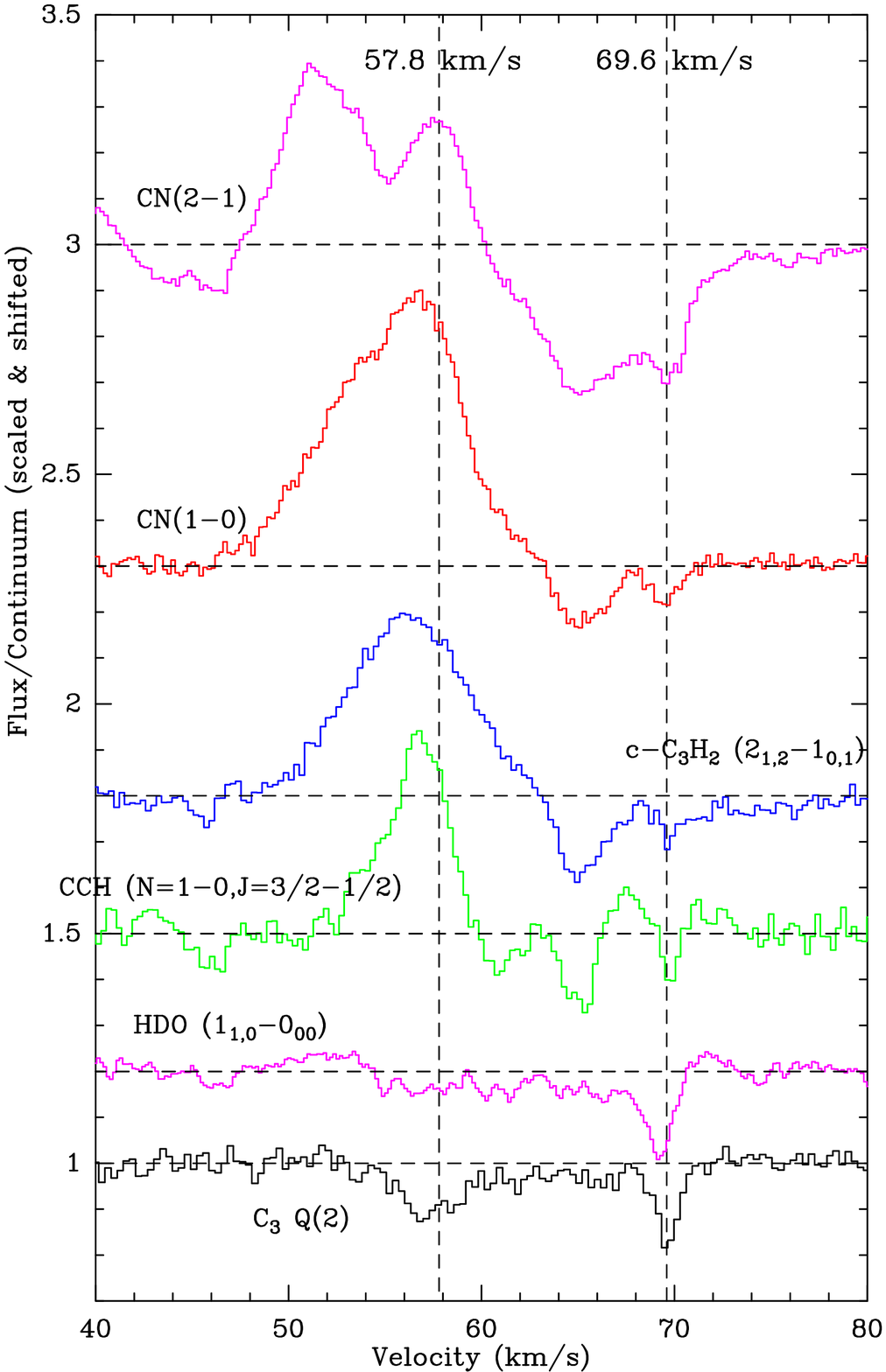}
\includegraphics[width=0.45\textwidth]{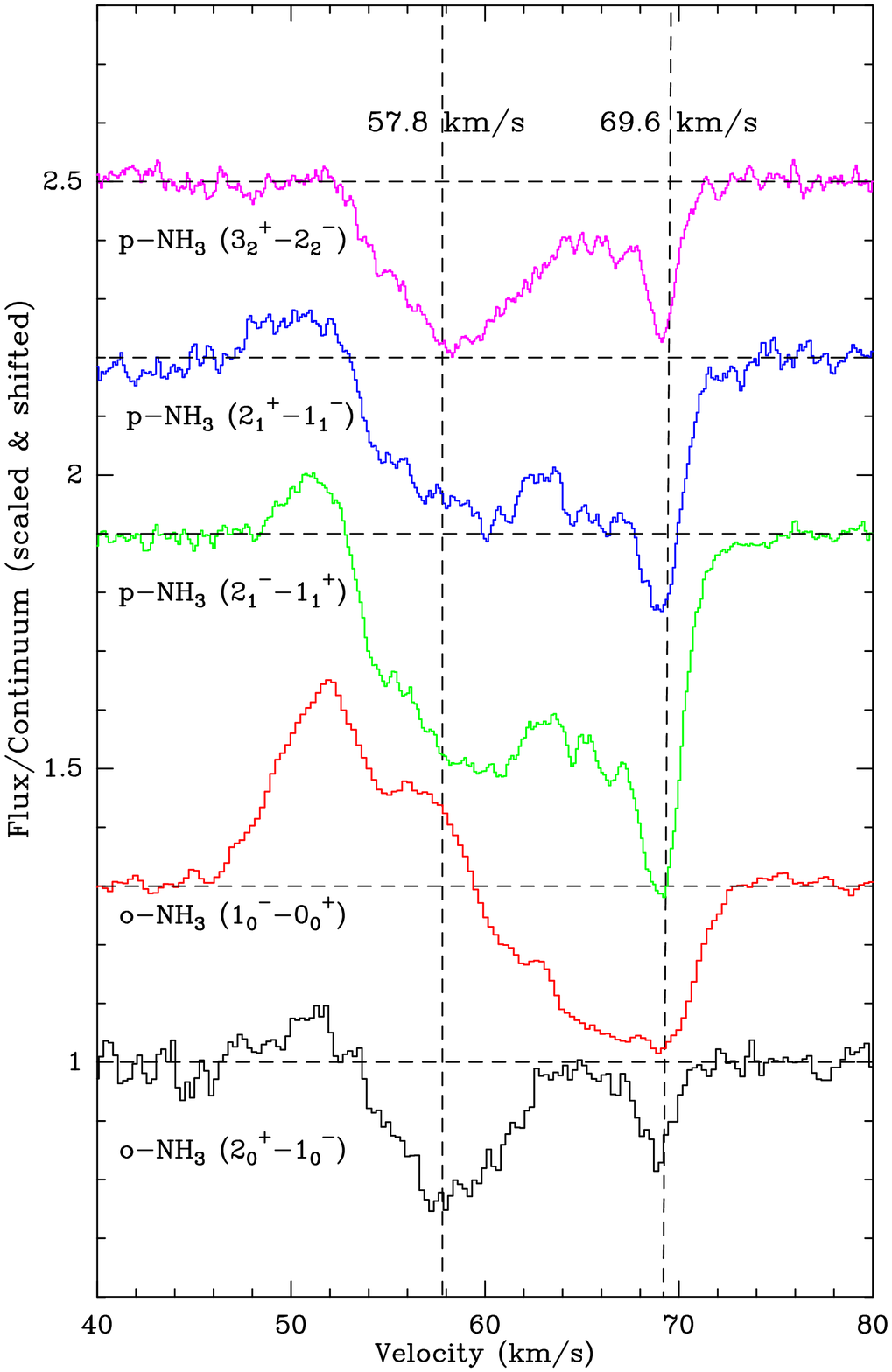}
\caption{{\bf Left:} comparison of velocity structure seen in 
\cthree\ spectra with that seen in transitions of CCH, \cthreehtwo,
 HDO, and CN. The H$_2^{18}$O spectrum along with a Gaussian fit are taken from
\citet{flagey2013}. {\bf Right:} multiple transitions of NH$_3$ observed
with Herschel. The dashed vertical lines correspond to the 57.8 and
69.6\,\kms\ components as found in the \cthree\ observations (see
Table~\ref{tab_gaussfit}).
\label{fig_compspec}}
\end{center}
\end{figure*}

Figure~\ref{fig_c3spec} shows the observed \cthree\ spectra normalized
to the (single-sideband) continuum level.  We identify two major
absorption dips around 58 and 70~\kms\ in the spectra of all three
transitions of \cthree. Figure~\ref{fig_compspec} shows the observed
spectra of single or multiple transitions of CCH, c-C$_3$H$_2$, NH$_3$,
CN, HDO, and  H$_2^{18}$O along with the \cthree\ ($Q(2)$) spectrum for
comparison. The H$_2^{18}$O 1$_{1,1}$--0$_{0,0}$ spectrum was observed
with Herschel/HIFI as part of the PRISMAS program and has been presented
by \citet{flagey2013}.  

While the 58\,\kms\ component appears either in emission or in
absorption, depending on the transitions, the narrow 70~\kms\ feature
can be seen only as a deep absorption feature  for the different species
presented here.  The absorption dip at 58~\kms\ is broad and most likely
is a combination of the 57~\kms\ component associated with the \HII\
region W51e2 and the \HI\ cloud at 62~\kms.  We note that although the
68\,\kms\ feature associated with the filament has been relatively
easily detected in CO emission \citep{mufson1979,carpenter1998}, the
line profiles are much broader than the 70\,\kms\ absorption dip
detected here.  Based on the results of our analyses we show in the
following sections that the 70\,\kms\ absorption feature can only be
explained in terms of a high-density ($>$10$^5$\,\cmcub) clump that
formed within the filament.

\section{Estimations of the column and volume densities}

Tables~\ref{tab_basobs} and \ref{tab_gaussfit} summarize the basic
observational results for CCH, \cthreehtwo, NH$_3$, HDO, and \cthree.
The central velocities, linewidths, and main-beam temperatures of all
species are derived by fitting Gaussian components to the 70~\kms\
feature of the observed spectra. In the following analyses, we assumed
that the foreground cloud at 70\,\kms\ completely fills the beam.

\begin{table*}[]
\begin{center}
\caption{Basic observational results for NH$_3$, HDO, CCH, and
\cthreehtwo\ \label{tab_basobs}}
\begin{tabular}{cccccc}
\hline
\hline
Species &Transition & \Tc & \vlsr\ & $\Delta V$ & $\int{T_{\rm mb} d\rm V}$\\
& & K & \kms\ & \kms\ & K~\kms\ \\
\hline
ortho-NH$_3$ & 1$_0^-$--0$_0^+$ & 1.40$\pm$0.04 & \ldots                  &\ldots                 &\ldots\\
                       & 2$_0^+$--1$_0^-$ & 7.40$\pm$0.29 & 68.83$\pm$0.06 & 2.03$\pm$0.16 & 3.44$\pm$0.49\\
para-NH$_3$ & 2$^-_1$--1$^+_1$ & 7.30$\pm$0.11 & 68.79$\pm$0.02 & 3.14$\pm$0.07 & 15.13$\pm$0.49\\
                      & 2$^+_1$--1$^-_1$  & 7.40$\pm$0.29 & 68.78$\pm$0.04 & 3.30$\pm$0.10 & 15.62$\pm$0.92\\
                      & 3$_2^+$--2$_2^-$  & 9.80$\pm$0.21 & 69.07$\pm$0.02 & 2.08$\pm$0.05 & 6.23$\pm$0.24\\
HDO & 1$_{1,1}$--0$_{0,0}$ & 7.10$\pm$0.10 & 69.17$\pm$0.02 & 1.52$\pm$0.06 & 1.64$\pm$0.10\\

CCH & 1$_{3/2,1}$--0$_{1/2,1}$& 1.70$\pm$0.09 & 56.69$\pm$0.06 & 3.50$\pm$0.50 & 5.40$\pm$0.92 $\star$\\
& & 1.70$\pm$0.09 & 60.76$\pm$0.14 & 1.30$\pm$0.20 & 0.53$\pm$0.18\\
& & 1.70$\pm$0.09 & 65.01$\pm$0.09 & 1.50$\pm$0.20 & 0.99$\pm$0.23\\
& & 1.70$\pm$0.09 & 69.79$\pm$0.09 & 0.68$\pm$0.21 & 0.30$\pm$0.17\\
\cthreehtwo & 2$_{1,2}$--1$_{0,1}$ & 1.00$\pm$0.04 & 56.17$\pm$0.15 & 6.75$\pm$0.36 & 4.77$\pm$0.09 $\star$\\
& & 1.00$\pm$0.04 & 65.26$\pm$0.05 & 2.63$\pm$0.12 & 0.78$\pm$0.09\\
& & 1.00$\pm$0.04 & 69.70$\pm$0.15 & 0.54$\pm$0.20 & 0.13$\pm$0.09\\
\hline
\end{tabular}
\end{center}

$\star$ Lines are detected in emission.
\end{table*}

\subsection{\cthree\ LTE modeling}

We fitted the three detected transitions of \cthree\ simultaneously
using two Gaussian components corresponding to two velocity components
per transition. The fit derives a common \vlsr\ and full width at half
maximum ($\Delta v$) for each of the velocity components for all
transitions as well as the strength of the absorption dips.
Table~\ref{tab_gaussfit} presents the single-sideband continuum levels
(estimated from double-sideband temperatures assuming a sideband gain
ratio of unity) and the opacities derived from the Gaussian fitting.

\begin{table}[h]
\begin{center}
\caption{Parameters derived for \cthree\ by simultaneous Gaussian
fitting of all observed transitions $^{a}$.
\label{tab_gaussfit}}
\begin{tabular}{lccrr}
\hline
\hline
Transition &\Tc$^{b}$ & \vlsr\ & $\int{\tau d\rm v}$ & $N_{\rm l}$\\
& \hspace*{0.3cm}[K] & [\kms] &  [\kms] &  [10$^{13}$\cmsq] \\
\hline
\noalign{\smallskip}
$P(4)$ & 10.5$\pm$1.3 & 57.8 & 0.14$\pm 0.05$ & 3.5$\pm1.0$  \\
       &              & 69.6 & 0.11$\pm 0.02$ & 2.7$\pm0.5$  \\
&&&&\\
$Q(2)$ & 10.5$\pm$0.6 & 57.8 & 0.30$\pm0.05$ & 2.5$\pm0.3$\\
       &              & 69.6 & 0.31$\pm 0.03$ & 2.6$\pm0.2$\\
&&&&\\
$Q(4)$ & 10.7$\pm0.7$ & 57.8 & 0.49$\pm 0.07$ & 4.1$\pm0.4$ \\
       &              & 69.6 & 0.23$\pm 0.02$ & 1.9$\pm0.2$ \\
\hline
\end{tabular}
\end{center}
$^{a}$ The fitted velocity components are  component 1: \vlsr =
57.81$\pm0.01$~\kms, $\Delta V$ = 4.72$\pm0.38$~\kms, component 2:
\vlsr = 69.55$\pm0.04$~\kms, $\Delta V$ = 1.41$\pm0.10$~\kms.\\
$^{b}$ Single sideband continuum main beam temperatures ($\eta_{\rm mb}$ = 0.69)
\end{table}

Assuming LTE we used the level-specific column densities derived from
the \cthree\ absorption dips (Table~\ref{tab_gaussfit}) following the
rotation diagram method.  For the 58~\kms\ component we estimate the
rotation temperatures (\Trot) and total column density ($N$(\cthree)) to
be 10\,K and 1.0$\times 10^{14}$~\cmsq, respectively, for the 70~\kms\
component these are 10\,K and 6.0$\times 10^{13}$~\cmsq, respectively.
Since the determination of \Trot\ and the column density involves
fitting of only two datapoints with associated uncertainties to estimate
two parameters, the derived parameters are expected to have very large
uncertainties. Additionally, since the far-infrared continuum radiation
is known to contribute significantly to the excitation of \cthree\
\citep{roueff2002}, an accurate estimate of temperature can only be
obtained using rigorous radiative transfer treatment. We therefore only
used the derived total column densities from the LTE analysis.

\subsection{Column density estimates from LTE modeling: CCH and
\cthreehtwo}

We used the IRAM-30m \cthreehtwo\ observations presented by
\citet{gerin2011} and the IRAM-PdBI CCH observations to compute the
column densities in the absorption components along the line of sight to
W51e2 using the CASSIS software\footnote{A software developed by
IRAP-UPS-CNRS: http://cassis.irap.omp.eu}.

\begin{table}[h]
\begin{center}
\caption{Derived parameters for the absorption components detected in CCH and
\cthreehtwo\ spectra
\label{tab_cch_c3h2}}
\begin{tabular}{ccccc}
\hline
\hline
Species & \vlsr\ & $\Delta$ v & $\tau$ & N \\
& \kms\ & \kms\ &   & \cmsq \\
\hline
CCH & 69.8 & 0.7 & 0.37 & 4.9~10$^{13}$ \\
& 65.0 & 1.5 & 0.52 & 1.3~10$^{14}$ \\
& 60.8 & 1.3 & 0.28 & 6.0~10$^{13}$ \\
\cthreehtwo\ & 69.7 & 0.54 & 0.23 & 7.5~10$^{11}$ \\
& 65.3 & 2.63 & 0.35 & 5.5~10$^{12}$ \\
\hline
\hline
\end{tabular}
\end{center}
\end{table}

For CCH we used the average (taken over 5\arcsec) spectrum centered on
the HIFI position observed with PdBI for modeling.  We detected the
70\,\kms\ absorption feature in only one transition of each of CCH and
\cthreehtwo.  Since  both CCH and \cthreehtwo\ lines at 70\,\kms\ appear
in emission instead of absorption for \Tex $>$3\,K,  we derived the
molecular column densities for each velocity component in absorption
assuming that the excitation temperature equals the CMB.  The ratio of
the column densities of CCH and c-C$_3$H$_2$ in the 70\,\kms\ component
($\sim$ 67) is similar to the value found in the extended ridge
component (N(H$_2$) = $3\times 10^{23}$ cm$^{-2}$) of the OMC-1 region
\citep{blake1987} (see Sec. 7.1).

\subsection{Volume density estimates from NH$_3$}

\begin{table}[h]
\caption{Non-LTE ({\tt RADEX}) modeling of NH$_3$ and CN. \label{tab_nh3_cn}}
\begin{tabular}{cccrcr}
\hline
\hline
\noalign{\smallskip}
Species & \vlsr\ & $\Delta v$ & T$_{\rm kin}$ & $N$ & $n$(H$_2$)\\
        & \kms\ & \kms  &  K & cm$^{-2}$ & \cmcub\\
\hline
\noalign{\smallskip}
CN           &  69.5 & 0.6 &  6.0 & 5.5$\times$10$^{13}$ & 10$^5$\\
&&&&&\\
NH$_3$       &  68.8 & 2.0 & 30.0 & 8.4$\times$10$^{13}$ & 5$\times$10$^5$\\
\hline
\hline
\end{tabular}

The H$_2$ density and kinetic temperature found are valid to within 50\%.
\end{table}

Because the ortho--NH$_3$ fundamental transition at 572.5\,GHz are
highly optically thick, with blending in the line of sight (see
Fig.\,\ref{fig_compspec}), the 1168.5, 1214.8, 1215.2, and 1763.8 GHz
transitions were used to separate the different components and extract
the 69.2 $\pm$ 0.2 km~s$^{-1}$ component. The results from the Gaussian
fitting procedure are presented in Table \ref{tab_basobs}.  Indeed, the
572\,GHz transition  is affected by blending and saturation effects and
does not allow us to obtain a reliable estimate for the ortho-to-para
ratio in the 70\,\kms\ cloud, in addition to n${(\rm H_2)}$ and $T_{\rm
kin}$. Furthermore, we do not have additional observations that might be
used to identify the NH$_3$ formation process (temperature and mechanis
in particular) \citep{persson2012} and the H$_2$ ortho-para ratio
\citep{faure2013}, both of which are known to result in an NH$_3$
ortho-para ratio significantly different from unity.  We assumed the
ortho-to-para ratio to be equal to the statistical equilibrium value of
unity. The ortho-- and para--NH$_3$ collisional coefficients with
para--H$_2$ \citep{danby1988} were used in the non-LTE {\tt RADEX}
computations within CASSIS. In the CASSIS database the ortho and para
separation is provided with partition functions for temperatures between
3 and 300 K.  Note that there is a discrepancy between the fitted line
widths of the two 2$_1$--1$_1$ ($\sim$ 3 km~s$^{-1}$) transitions and
the 2$_0^+$--1$_0^-$ and 3$_2^+$--2$_2^-$ (2.0 km~s$^{-1}$) transitions.
However, we argue that the line widths of the 2$_1$--1$_1$ transitions
are more affected by additional components in its wings than the more
isolated velocity components of the 3$_2^+$--2$_2^-$ and
2$_0^+$--1$_0^-$ transitions.  We therefore varied the line width around
a value of 2 km~s$^{-1}$ in the modeling. The best-fitting model for the
70\,\kms\ cloud is obtained for a line width of 2 km~s$^{-1}$ with an
H$_2$ density of 5$\times$ 10$^5$\,\cmcub, and a kinetic temperature of
$\sim 30$\,K, with a column density of 4.2 $\times$ 10$^{13}$~\cmsq\ for
each symmetry.  Hence the total estimated column density of NH$_3$ is
8.4$\times$10$^{13}$\,\cmsq.

\subsection{Volume density estimates from CN}

\begin{figure*} \begin{center}
\includegraphics[width=0.45\textwidth]{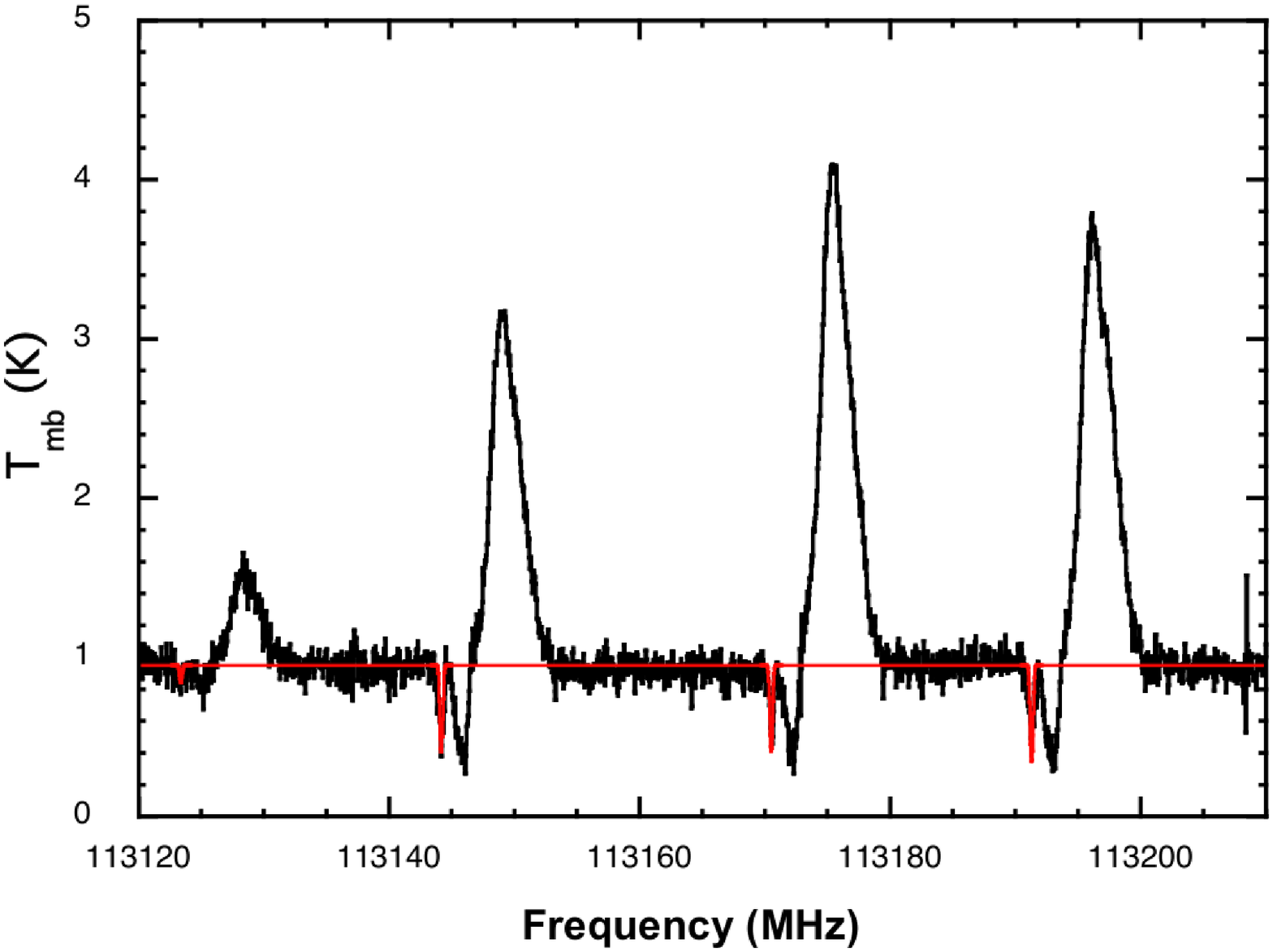}
\includegraphics[width=0.45\textwidth]{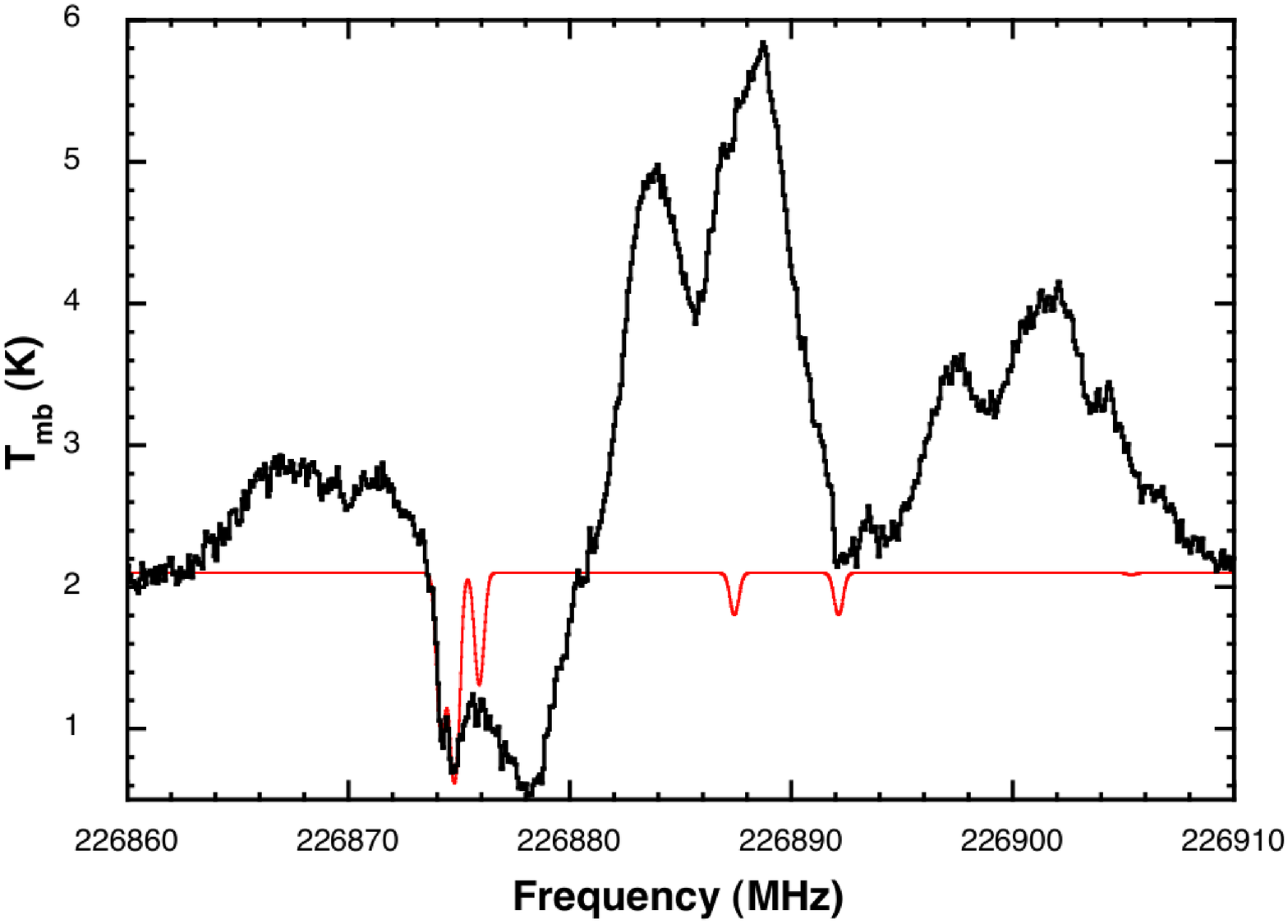}
\includegraphics[width=0.95\textwidth]{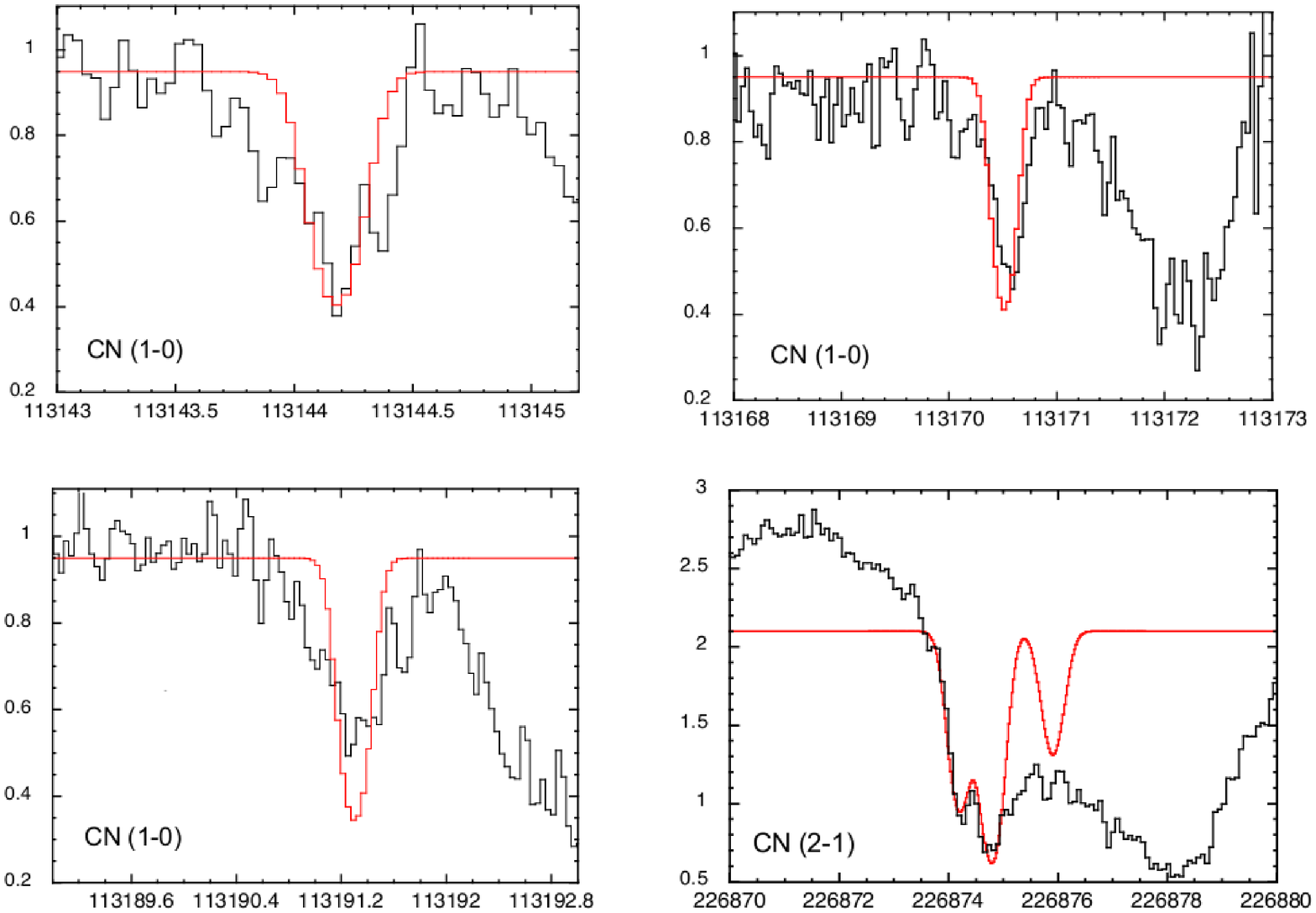} \caption{Top
panels: observed spectra of CN 1--0 ({\em left}) and CN 2--1 ({\em right})
shown in black. We also show the best-fit LTE model (using the CASSIS
software) corresponding to \Tex=3\,K, \vlsr = 69.5\,\kms, $\Delta v$=
0.6\,\kms, and $N_{\rm CN}$ = 5.8$\times$10$^{13}$\,\cmsq. The data and the
relevant best-fit model spectra are enlarged in the four panels below.
The details of the CN(1--0) transitions are middle row {\em left:}
1,1/2,1/2--0,1/2,3/2 and {\em right:} 1,1/2,3/2--0,1/2,1/2; bottom row
{\em left:} 1,1/2,3/2--0,1/2,3/2.
\label{fig_cn}} 
\end{center} 
\end{figure*}

We observed the $J$=1--0 and 2--1 transitions of CN using the IRAM
30\,m. CN 1--0 is a four component multiplet that appears in absorption
against the 0.95\,K continuum, and CN 2--1 is a six component multiplet,
four components of which appear in absorption against the 2.1\,K
continuum (the other components are blended with the emission components
at the source velocity).  Figure~\ref{fig_cn} shows the observed spectra
along with the best-fit  LTE model for \Tex=3\,K, \vlsr = 69.5\,\kms,
line width of 0.6\,\kms\ and a total column density of 5.8$\times
10^{13}$\,\cmsq.  Recently, the collision rates have been computed with
p-H$_2$ taking into account all the CN hyperfine levels
\citep{kalugina2012}.  Assuming the line width of the 70\,\kms\
component to be 0.6\,\kms, we ran a non-LTE model ({\tt RADEX}) to
reproduce the seven CN hyperfine components seen in absorption and found
a density of $n(\rm H_2)\sim 10^5$\,\cmcub (assuming that H$_2$ is
mainly in its para form) and a kinetic temperature of $\sim$ 6\,K with a
column density of $\sim$5.5$\times 10^{13}$\,\cmsq. Both the density and
temperature required to explain the observed CN spectra are lower than
the values required to explain the observed ammonia transitions (see
Table \ref{tab_nh3_cn}).  A possible reason for the low temperature
derived can be the possible dilution of the continuum within the beam,
combined with a cold foreground component that is covering part of the
continuum.  Assuming a higher continuum value (e.g. 1.7\,K, as found for
CCH) would lead to a 10 K kinetic temperature for the 70\,\kms\
component. We therefore consider 6\,K to be a lower limit for the
temperature derived from CN. However, given the significantly different
temperatures (30\,K) and densities (5$\times 10^5$\,\cmcub) found for
NH$_3$ along with the extremely narrow line width of the CN lines
compared with those found in NH$_3$, \cthree, and HDO, it is very likely
that CN arises from a colder and less dense envelope, possibly giving
rise to CCH and \cthreehtwo\ as well.

\subsection{HDO}

Most of the compact \HII\ regions observed within the PRISMAS Key
Program present a broad absorption profile in the 893\,GHz fundamental
transition of HDO at the source \vlsr. The W51 region is the only
high-mass source where this transition appears as a deep narrow ($<$ 2
km~s$^{-1}$) feature at a velocity higher than the source velocity. The
other HDO fundamental transition at 464\,GHz has also been observed but
the high-velocity absorption feature is not detected reliably above the
noise of the spectrum (rms = 0.13\,K).  From the narrow component
observed at 893~GHz, we can state that this component is clearly
localized and probably not extended along the line of sight.  Both HDO
ground-state transitions have high critical densities ($>$ 10$^7$
cm$^{-3}$) and do not provide strong constraints on the kinetic
temperature or density for the range found with the NH$_3$ and CN
analysis. Because the NH$_3$ transitions trace a dense region in the
filament, which is also more likely to emit in HDO than the lower
density and lower temperature region that is traced by CN, we estimate
for $T_{\rm kin}$ = 30\,K and $n(H_2)$ = 5$\times 10^5$\,\cmcub\ the
total HDO column density under non-LTE to be 1.5$\times 10^{12}$
cm$^{-2}$, which is similar to the LTE value with an excitation
temperature of 3\,K.

\section{Chemical modeling of abundance}
\label{sec-chemmod}

In the absence of a direct estimate for the H$_2$ column density for the
70\,\kms\ clump, we expressed the abundances of all the observed
molecules \cthree, \cthreehtwo, NH$_3$, CN, and HDO  (Table
~\ref{tab_obsabund}) relative to the column density of CCH and compared
this with the results of the Ohio State University (OSU) gas-grain
chemical model with a warm-up \citep{hhl92,gh06}.  A previous version of
this model has already been used to interpret the abundance of \cthree\
in the warm envelopes of hot cores such as W31C, W49N, and DR21OH
\citep{mookerjea2010,mookerjea2012}.  While the 70~\kms\ component is
not directly associated to the hot core, the necessity of grain warm-up
in explaining the abundance of \cthree\ in dense environments was
illustrated by {\citet{hassel2011}}. As explained by
\citet{mookerjea2012}, formation of \cthree\ in dense environments
cannot proceed via the gas-phase pathway as in diffuse clouds and
requires desorption (either by radiation or by cosmic ray) of CH$_4$
from icy grain mantles to produce C$_2$H$_2$, which reacts further to
produce \cthree.

The chemical model is adapted from models of the hot-core phase of
protostellar sources \citep{viti04,gh06}. Standard hot-core chemistry
requires that dust grains build up icy mantles at early times, when the
temperature is low ($\sim$10 K) and the core is in a state of collapse
\citep{brown1988}.  Later, the formation of a nearby protostar quickly
warms the gas and dust (warm-up phase), re-injecting the grain mantle
material into the gas phase, and stimulating a rich chemistry. Thus,
hot-core models can involve three distinct phases: (i) the collapse of a
cold clump of interstellar gas to high densities, (ii) heating of this
clump, and the subsequent evaporation of grain mantles, and (iii) the
hot-core phase.  The present model examines (ii), the warm-up phase, and
starts from the time at which the clump has collapsed to the high
density and considers heating of the constant density clump and
subsequent grain surface and gas-phase chemistry, corresponding to Phase
2 of \citet{viti04}, based on the approximation of \citet{hhg08}.  In
the warm-up approach, a one-point parcel of material undergoes an
initially cold period of $T_{0}=10$~K with a duration of $t=10^5$ yr,
followed by a gradual temperature increase to a maximum temperature,
$T_{\rm max}$.   A heating time-scale of $t_{\rm{h}}=0.2$~Myr was
adopted following \citet{gh06}, in which the temperature reaches $T_{\rm
max}$ by 0.3 Myr.  In the warm-up, the dust and gas temperatures are
assumed to be equal and follow the same evolution.  Unlike the hot-core
models, it is unclear whether the filament harbors star formation. We
therefore considered values of $T_{\rm max}$ that represent relatively
mild warming.  The physical parameters adopted are the same as used by
\citet{gwh07} and \citet[Table 1;][]{hhg08}, including a canonical value
for cosmic-ray ionization rate $\zeta=1.3 \times 10^{-17}$\,s$^{-1}$.
The simplified warm-up and the estimates for the parameters were
previously adopted by {\citet{hhg08}} following {\citet{gh06} to
investigate the details of chemical evolution without a strict
dependence on a particular dynamical model, and have since been used in
{\citet{hassel2011}} and {\citet{mookerjea2012}}.  Although the dense
clump may not contain a hot core, we adopted the same approach in this
model for the same general reason, and in particular to incorporate
reactions involving D-bearing species into a familiar modeling
procedure.}

The gas-grain chemistry reaction network has been expanded since the
previous C$_3$ models to include HDO, and now comprises 7882 reactions
involving 735 gaseous and surface species.  We included the major
gaseous D-exchange reactions of the UMIST database \citep{umist12}, the
major surface reactions identified by \citet{stantcheva03}, and
gas-phase reactions related to HDO using the approach of
\citet{aikawa12}.  The initial chemical abundances were adopted from
\citet{gwh07}, and include the initial abundance [HD] $ =
1.5\times10^{-5}\times [\rm{H_2}]$ \citep{osamura05}.

\begin{figure}
\begin{center}
\hspace*{-0.5cm}
\includegraphics[width=0.48\textwidth]{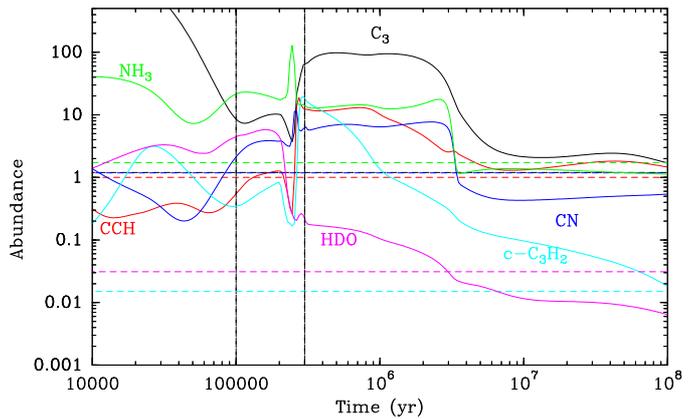}
\caption{Calculated abundances of different species relative to the
observed column density of CCH, as a function of time for a constant
density $n(\rm H_2)$ = 5$\times 10^5$~\cmcub\ and a warm-up to
\Tmax=30\,K from 10 K, where the warm-up and other conditions are
described in \S~\ref{sec-chemmod}.  The dashed horizontal lines
correspond to the observed abundances and are color-coded with the
calculated abundances. The vertical lines at $t=0.1$ and 0.3\,Myr
indicate the onset and end of warm-up of grains, respectively.
\label{fig_abund}}
\end{center}
\end{figure}

The observed multiple transitions of NH$_3$ provide the only independent
estimate of the density and temperature of the filament that gives rise
to the 70\,\kms\ component. Here we present a comparison of the observed
abundances for all molecules considered with the abundances predicted by
the chemical model for which $T_{\rm max}$ is 30\,K and the molecular
hydrogen density is 5$\times 10^5$\,\cmcub\ (Fig.\,\ref{fig_abund}).
The predicted abundances of all molecules are significantly different
from the observed values at times earlier than $\sim 3$\,Myr.  The
abundances of NH$_3$, CCH, and CN predicted by the model match the
observed abundances well at all times beyond $t$ = 3.5 Myr. The model
abundance of \cthreehtwo, is significantly higher than the observed
abundance, at all times.  Previous models using a similar chemical
network have also demonstrated a larger calculated abundance than the
observation \citep[Figure 4;][]{hassel2011}. It is also possible that
\cthreehtwo\ arises in a colder and less dense envelope (similar to CN),
than the values of $n({\rm H_2})$ and $T_{\rm max}$ used in the model
presented here.  Beyond 5\,Myr the model abundance of \cthree\
approaches the observed abundance for all times up to $t$=70\,Myr. For
\cthree\ the model abundance is within a factor of 3 of the observed
abundance for all times later than 5\,Myr. The HDO model abundance is
consistent to within a factor of 2.5 of the observed abundance between
3.5 to 30\,Myr and diverges farther at later times.

\begin{table}[h]
\begin{center}
\caption{Abundances of all chemical species estimated relative to the observed 
CCH column density in the 70\,\kms\ clump.
\label{tab_obsabund}}
\begin{tabular}{llr}
\hline
\hline
Species &  \multicolumn{2}{c}{70\,\kms\ clump}\\
& $N$ & $X^{a}$ \\
& 10$^{13}$\,cm$^{-2}$  & \\
\hline
\cthree\  & 6.0 & 1.2  \\
CCH   & 4.9 & 1.0 \\
\cthreehtwo\ &  0.08 & 0.016\\
NH$_3$ &  8.4 & 1.72  \\
CN &  5.5 & 1.12   \\
HDO &  0.15 & 0.031   \\
\hline
\end{tabular}
\end{center}

$X^{a}$: Abundance relative to $N$(CCH) = 4.9$\times 10^{13}$\,\cmsq.
$^{b}$ \citet{mookerjea2012}
\\
\end{table}

\section{Discussion}

In the direction of W51e2 we have for the first time detected HDO and
\cthree\ in  a dense molecular clump at 70\,\kms\ that is not directly
associated with hot star-forming cores. 

\cthree\ in dense clouds has so far only been detected in the warm
envelopes associated with star-forming cores. Abundance of \cthree\ in
dense clouds is explained via warming-up of grain mantles to release
CH$_4$ and subsequently C$_3$H$_2$, the first step in the chemical
pathway for the formation of \cthree, as explained in detail in
{\citet{hassel2011}} and {\citet{mookerjea2012}}. Thus, detection of
\cthree\ in the dense clump at 70\,\kms\ otherwise not identified as a
star-forming region is as surprising as the detection of deuterated
water in the cloud that is physically not related to the W51 region. HDO
is most likely formed in the ice mantles within cold ($<$20\,K) translucent
clouds \citep{cazaux2011}  in regions shielded from UV radiation.
Indeed, at $A_{\rm v} \gtrsim 3$ \citep{whittet1988}, species from the
gas phase accrete onto dust grains and initiate ice formation.  
After the formation of water and deuterated
water in the ice mantles in the
star formation scenario, 
the environment undergoes gravitational
collapse  and the local density increases, leading to the formation of a
protostar.  The surrounding medium then heats up and releases the icy
mantles into the gas phase, leading to a significant HDO abundance in
the gas phase \citep{coutens2012,coutens2013b}. Based on the non-LTE
modeling of the five observed NH$_3$ transitions, we characterize the
region of the that filament gives rise to the 70\,\kms\ feature as having
$T_{\rm kin}$ = 30\,K and $n({\rm H_2})$ = 5$\times 10^5$\,\cmcub. The
abundances predicted by the grain warm-up chemical model reproduce the
observed abundances to within factors of 2 to 3 at all times later than
3.5 Myr for all species except for \cthreehtwo.  


\subsection{CCH/\cthreehtwo\ ratio in the clump}

Previous observations of CCH and \cthreehtwo\ toward dense dark clouds
and Photon Dominated Regions (PDRs) derived a tight correlation between
the two species.  The ratio $N$(CCH)/$N$(\cthreehtwo) is $\sim 61$ in
OMC-1 \citep{blake1987} and it ranges between 10 and 25 in PDRs such as
Horsehead nebula and IC63 whereas in dark clouds such as TMC-1, and L134N
the ratio is $\sim 1$ \citep{teyssier2004}. For the dense clump detected
toward W51e2  the observed $N$(CCH)/$N$(\cthreehtwo) of 67 is closer to
the value observed in the dense active star-forming region OMC-1 than
to the values seen in UV-illuminated clouds and dark clouds.

\subsection{Deuterium enrichment in the clump}

Using Herschel/HIFI, a number of transitions of H$_2$O and H$_2$$^{18}$O
species corresponding to a range of excitations have been observed for
the first time. The 70\,\kms\ feature is detected unambiguously in the
H$_2$O 2$_{0,2}$--1$_{1,1}$ transition as well as the H$_2$$^{18}$O
1$_{1,1}$--0$_{0,0}$ transition \citep{flagey2013}.  However, an
accurate determination of the total H$_2$O column density is compromised
by the crowded environment along the line of sight with high optical
depth, as well as by partial blending of the components. The column
density for the less abundant para-H$_2$$^{18}$O species is estimated to
be (7 $\pm$ 1) 10$^{11}$ cm$^{-2}$ for the 70\,\kms. The abundance ratio
between $^{16}$O and $^{18}$O is commonly accepted to be $\sim$ 560
\citep{wilson1994}, although significant variation has been observed
throughout the Galaxy.  Accordingly, we estimate a D/H ratio in water
vapor of about 9.6 $\times$ 10$^{-4}$ (assuming an ortho-para ratio of
3), similar to the value found in Orion\,KL by \citet{neill2013}.  This
ratio has also been estimated from HDO observations of the  low-mass
protostar IRAS 16293-2422 \citep{coutens2012,coutens2013}. The derived
values of HDO/H$_2$O were  1.8$\times 10^{-2}$ in the hot core, 5$\times
10^{-3}$ in the cold envelope, and about 4.8$\times 10^{-2}$ in the
external photodesorption layer. The value estimated from our analysis is
consistent with the value found in the cold envelope of the IRAS
16293-2422 low-mass star-forming region, in which the ices are not
affected by thermal desorption or photo-desorption by the FUV field.
For the best-fit grain warm-up model that reproduces the observed
abundances of the other molecules, at times later than 3.5 Myr the
HDO/H$_2$O ratio drops from 0.002 to 0.001 for an initial HD/H$_2$ ratio
of $1.5\times 10^{-5}$. The calculated HDO/H$_2$O ratio is consistent
with the observed ratio of 9.6 $\times$ 10$^{-4}$. Although HDO forms in
the medium density, pre-collapse phase, it mostly remains on grain
mantles, since the nonthermal desorption mechanisms have a limited
efficiency.  In these models, which reach only 30\,K, the release of HDO
into the gas phase is attributed to nonthermal processes related to
cosmic-ray interactions rather than warming up or direct FUV
photodesorption.  Although the chemical model predicts an HDO/H$_2$O
ratio consistent with the observed value for $t>3.5$\,Myr, it
overestimates the HDO/H$_2$O ratio by more than a factor of 10 for
earlier times $t<1$\,Myr. The calculated HDO/H$_2$O ratio is
substantially enriched above the initial HD/H$_2$ ratio, primarily
because of the grain surface formation route of HDO. The ratio is
furthermore enriched above the observed value because of the lower
column density of H$_2$O calculated by the chemical models.  We explored
the possibility of enhanced cosmic-ray ionization rates. For enhanced
cosmic ray rates the chemical models produce abundances very different
from the observed values for all molecules.  This result for the
70\,\kms\ cloud is in contrast to the higher values of $\zeta$ deduced
by \citet{indriolo2012} along the same line of sight for the absorption
features arising because of diffuse foreground.   This implies that the
absorbing gas is not diffuse, but is dense with a standard value of
$\zeta$.

\subsection{Possible origin of the clump}

The 70\,\kms\ absorption dip caused by the foreground material
in the direction of W51e2, although not associated with a star-forming
core, shows chemical abundances typically found in the envelopes of
low-mass protostellar candidates.   We propose that the feature arises
in a dense ($n$(H$_2$)=(1-5)$\times 10^5$\,\cmcub) and cold (10-30\,K)
clump that is formed within the much larger scale filament (detected in
CO) deemed to be interacting with the W51 main molecular cloud. A
possible scenario for the formation of the dense clump can be outlined
based on the collision of the filament with W51 main \citep{kang2010}.
In this scenario, cloud-cloud collision lead to the compression of the
interface region and initiate the formation of stars as seen in W51
\citep{habe1992}.  The molecular clumps at the interface are heated, but
the molecular clumps on the trailing side remain cold and hence appear
in absorption.  Models for collision between two different clouds
suggest that such collisions disrupt the larger cloud while the small
cloud is compressed and subsequently forms stars. It is possible that
the dense clump detected in absorption is formed on the trailing side of
the filament that interacts with the main molecular cloud. Based on its
density and somewhat elevated temperatures ($\sim 30$\,K), it is also
possible that this cloud is also collapsing to eventually form stars.
In the absence of any observational evidence, we currently do not 
consider the source to be an IRDC. We suggest that the nondetection of
this star-forming clump in the continuum is due to its chance
coincidence along the line of sight with the much stronger W51e2
continuum source. 

Higher spectral and angular resolution observations of the foreground
gas responsible for the absorption dip at $\sim 70$\,\kms\ are needed to
understand its spatial distribution. We also propose to explore a
clearer detection of the protostar using mid-infrared observations of
the strong spectral features of molecular ices toward this region.

{}

\acknowledgement 

We thank the anonymous referee for the suggestions that helped improve
the clarity of the paper significantly.  HIFI has been designed and
built by a consortium of institutes and university departments from
across Europe, Canada and the United States under the leadership of SRON
Netherlands Institute for Space Research, Groningen, The Netherlands and
with major contributions from Germany, France and the US. Consortium
members are: Canada: CSA, U.~Waterloo; France: CESR, LAB, LERMA, IRAM;
Germany: KOSMA, MPIfR, MPS; Ireland, NUI Maynooth; Italy: ASI,
IFSI-INAF, Osservatorio Astrofisico di Arcetri-INAF; Netherlands: SRON,
TUD; Poland: CAMK, CBK; Spain: Observatorio Astron\'omico Nacional
(IGN), Centro de Astrobiolog\'a (CSIC-INTA).  Sweden: Chalmers
University of Technology - MC2, RSS \& GARD; Onsala Space Observatory;
Swedish National Space Board, Stockholm University - Stockholm
Observatory; Switzerland: ETH Zurich, FHNW; USA: Caltech, JPL, NHSC.  G.
Hassel gratefully acknowledges helpful discussions and rate information
provided by Yuri Aikawa. C. Vastel is grateful to F. Lique for providing
the CN collisional rates in the LAMDA database format.  T. Harrison
thanks the Summer Scholars grant from the Center for Undergraduate Research
and Creative Activity (CURCA), Siena College.  This paper has made
extensive use of  the SIMBAD database, operated at CDS, Strasbourg,
France. This publication makes use of molecular line data from the
Boston University-FCRAO Galactic Ring Survey (GRS). The GRS is a joint
project of Boston University and Five College Radio Astronomy
Observatory, funded by the National Science Foundation under grants
AST-9800334, AST-0098562, AST-0100793, AST-0228993, \& AST-0507657.
This paper is partly based on observations obtained with the IRAM
Plateau de Bure interferometer and 30~m telescope. We are grateful to
the IRAM staff at Plateau de Bure, Grenoble for their support during the
observations and data reductions.  IRAM is supported by INSU/CNRS
(France), MPG (Germany), and IGN (Spain).  Part of the research was
carried out at the Jet Propulsion Laboratory, California Institute of
Technology, under a contract with the National Aeronautics and Space
Administration. 


\begin{thebibliography}{}

\bibitem[Aikawa et al.(2012)]{aikawa12} Aikawa, Y., Wakelam, V.,
  Hersant, F., Garrod, R.T., \& Herbst, E. 2012, \apj, 760, 40

\bibitem[Andr{\'e} et al.(2010)]{andre2010} Andr{\'e}, P., Men'shchikov,
A., Bontemps, S., et al.\ 2010, \aap, 518, L102

\bibitem[Arnal \& Goss(1985)]{arnal1985} Arnal, E.~M., \& Goss, W.~M.\ 
1985, \aap, 145, 369


\bibitem[Bally et al.(1987)]{bally1987} Bally, J., Lanber, W.~D., Stark,
A.~A., \& Wilson, R.~W.\ 1987, \apjl, 312, L45

\bibitem[Bergin et al.(2010)]{bergin2010} Bergin, E.~A., Hogerheijde, M.~R., Brinch, C., et al.\ 2010, \aap, 521, L33

\bibitem[Bieging (1975)]{bieging1975} Bieging, J. 1975, in Lecture Notes
in Physics, 42, H II Regions and Related Topics, ed. T. L. Wilson \& D.
Downes (Berlin : Springer), 443

\bibitem[Blake et al.(1987)]{blake1987} Blake, G.~A., Sutton, 
E.~C., Masson, C.~R., \& Phillips, T.~G.\ 1987, \apj, 315, 621

\bibitem[Brown et al.(1988)]{brown1988} Brown, P.~D., Charnley, S.~B.,
\& Millar, T.~J.\ 1988, \mnras, 231, 40

\bibitem[Carpenter \& Sanders(1998)]{carpenter1998} {Carpenter}, J.~M.
and {Sanders}, D.~B.  1998, \aj, 116, 1856

\bibitem[Cazaux et al.(2011)]{cazaux2011} Cazaux, S., Caselli, P., 
\& Spaans, M.\ 2011, \apjl, 741, L34



\bibitem[Cheung et al.(1968)]{cheung1968} Cheung, A.~C., Rank, D.~M.,
Townes, C.~H., Thornton, D.~D., \& Welch, W.~J.\ 1968, Physical Review
Letters, 21, 1701

\bibitem[Coutens et al.(2012)]{coutens2012} Coutens, A., Vastel, C.,
Caux, E., et al.\ 2012, \aap, 539, A132

\bibitem[Coutens et al.(2013a)]{coutens2013} Coutens, A., Vastel, 
C., Cazaux, S., et al.\ 2013a, \aap, 553, A75

\bibitem[Coutens et al.(2013b)]{coutens2013b} Coutens, A., Vastel, C.,
Cabrit, S., et al.\ 2013b, \aap, 560, A39

\bibitem[Danby et al.(1988)]{danby1988} Danby, G., Flower, D.~R., 
Valiron, P., Schilke, P., \& Walmsley, C.~M.\ 1988, \mnras, 235, 229

\bibitem[de Graauw et al.(2010)]{deGraauw2010} de Graauw, T., et al.\
2010, \aap, 518, L6 

\bibitem[Faure et al.(2013)]{faure2013} Faure, A., Hily-Blant, P., Le
Gal, R., Rist, C., \& Pineau des For{\^e}ts, G.\ 2013, \apjl, 770, L2

\bibitem[Flagey et al.(2013)]{flagey2013} Flagey, N., Goldsmith, P.~F.,
Lis, D.~C., et al.\ 2013, \apj, 762, 11 


\bibitem[Garrod \& Herbst(2006)]{gh06} Garrod, R.T., \& Herbst, E.
2006, \aap, 457, 927

\bibitem[Garrod et al.(2007)]{gwh07} Garrod, R.T., Wakelam, V., \&
Herbst, E. 2007, \aap, 467, 1103

\bibitem[Gaume et al.(1993)]{gaume1993} Gaume, R.~A., Johnston, 
K.~J., \& Wilson, T.~L.\ 1993, \apj, 417, 645




\bibitem[Gerin et al.(2011)]{gerin2011} Gerin, M., Ka{\'z}mierczak,
M., Jastrzebska, M., Falgarone, E., Hily-Blant, P., Godard, B., \& de
Luca, M.\ 2011, \aap, 525, A116 


\bibitem[Godard et al.(2010)]{godard2010} Godard, B., Falgarone, E., Gerin, M.,
Hily-Blant, P., \& de Luca, M.\ 2010, \aap, 520, A20 


\bibitem[Habe \& Ohta(1992)]{habe1992} Habe, A., \& Ohta, K.\ 1992,
\pasj, 44, 203


\bibitem[Hasegawa et al.(1992)]{hhl92} Hasegawa, T.I., Herbst, E., \&
Leung, C.M. 1992, \apjs, 82, 167

\bibitem[Hassel et al.(2008)]{hhg08} Hassel, G.E., Herbst, E., \&
Garrod, R.T. 2008, \apj, 681, 1385

\bibitem[Hassel et al.(2011)]{hassel2011} Hassel, G.E., Harada, N., \&
Herbst, E. 2011, \apj, 743, 182

\bibitem[Hily-Blant et al.(2010)]{hilyblant2010} Hily-Blant, P., Maret, S.,
Bacmann, A., et al.\ 2010, \aap, 521, L52

\bibitem[Indriolo et al.(2012)]{indriolo2012} Indriolo, N., Neufeld,
D.~A., Gerin, M., et al.\ 2012, \apj, 758, 83

\bibitem[Jacq et al.(1990)]{jacq1990} Jacq, T., Walmsley, C.~M., Henkel,
C., et al.\ 1990, \aap, 228, 447

\bibitem[Kalugina et al.(2012)]{kalugina2012} Kalugina, Y, Lique, F.,
Klos, J., \mnras 422, 812

\bibitem[Kang et al.(2010)]{kang2010} {Kang}, M. and {Bieging}, J.~H. and 
{Kulesa}, C.~A. and {Lee}, Y. and {Choi}, M. and {Peters}, W.~L., \apjs, 190, 58

\bibitem[Koo(1997)]{koo1997} Koo, B.-C.\ 1997, \apjs, 108, 489

\bibitem[Lique et al.(2010)]{lique2010}Lique, F. 2010, J. Chem. Phys. 132, 
044311

\bibitem[Liu et al.(2011)]{fcliu2011} Liu, F.-C., Parise, B., Kristensen, L., et al.\ 2011, \aap, 527, A19

\bibitem[Low et al.(1984)]{low1984} Low, F.~J., Young, E., Beintema,
D.~A., et al.\ 1984, \apjl, 278, L19

\bibitem[Lucas \& Liszt(2000)]{lucas2000} Lucas, R., \& Liszt, H.~S.\
2000, \aap, 358, 1069


\bibitem[McElroy et al.(2013)]{umist12} McElroy, D., Walsh, C.,
    Markwick, A.J., Cordiner, M.A., Smith, K. \& Millar, T.J. 2013,
    \aap, 550, 36

\bibitem[Mehringer(1994)]{mehringer1994} Mehringer, D.~M.\ 1994, 
\apjs, 91, 713

\bibitem[Molinari et al.(2010)]{molinari2010} Molinari, S., Swinyard,
B., Bally, J., et al.\ 2010, \aap, 518, L100

\bibitem[Mookerjea et al.(2010)]{mookerjea2010} Mookerjea, B., et al.\
2010, \aap, 521, L13 

\bibitem[Mookerjea et al.(2012)]{mookerjea2012} Mookerjea, B., Hassel,
G.~E., Gerin, M., et al.\ 2012, \aap, 546, A75



%

\bibitem[Mufson \& Liszt(1979)]{mufson1979} Mufson, S.~L., \& Liszt,
H.~S.\ 1979, \apj, 232, 451

\bibitem[Neill et al.(2013)]{neill2013} Neill, J.~L., Wang, S., 
Bergin, E.~A., et al.\ 2013, \apj, 770, 142






\bibitem[Osamura et al.(2005)]{osamura05} Osamura, Y., Roberts, H. \&
  Herbst, E. 2005, \apj, 621, 348

\bibitem[Ostriker(1964)]{ostriker1964} Ostriker, J.\ 1964, \apj, 140,
1056

\bibitem[Ott(2010)]{ott2010} Ott, S.\ 2010, Astronomical Data Analysis
Software and Systems XIX, 434, 139

\bibitem[Pardo et al.(2001)]{pardo2001} Pardo, J.~R., Cernicharo, J.,
Herpin, F., et al.\ 2001, \apj, 562, 799

\bibitem[Persson et al.(2007)]{persson2007} Persson, C.~M., Olofsson,
A.~O.~H., Koning, N., et al.\ 2007, \aap, 476, 807

\bibitem[Persson et al.(2010)]{persson2010} Persson, C.~M., Black,
J.~H., Cernicharo, J., et al.\ 2010, \aap, 521, L45

\bibitem[Persson et al.(2012)]{persson2012} Persson, C.~M., De Luca, M.,
Mookerjea, B., et al.\ 2012, \aap, 543, A145

\bibitem[Pety(2005)]{pety05b} Pety, J.\ 2005, SF2A-2005: 
Semaine de l'Astrophysique Francaise, 721

\bibitem[Pety et al.(2005)]{pety2005} Pety, J., Teyssier, D.,
Foss{\'e}, D., Gerin, M., et al.\ 2005, \aap, 435, 885

\bibitem[Pickett et al.(1998)]{pickett1998} {Pickett}, H.~M. and
{Poynter}, R.~L. and {Cohen}, E.~A. and {Delitsky}, M.~L. and {Pearson},
J.~C. and {M{\"u}ller}, H.~S.~P., \jqsrt, 60, 883

\bibitem[Pilbratt et al.(2010)]{pilbratt2010} Pilbratt, G.~L.,
Riedinger, J.~R., Passvogel, T., et al.\ 2010, \aap, 518, L1 



\bibitem[Radi et al.(1988)]{radi1988} Radi, P.~P., Bunn, T.~L., 
Kemper, P.~R., Molchan, M.~E., \& Bowers, M.~T.\ 1988, \jcp, 88, 2809 

\bibitem[Roelfsema et al.(2012)]{roelfsema2012} Roelfsema, P.~R.,
Helmich, F.~P., Teyssier, D.et al., 2012, \aap, 537, A17 

\bibitem[Roueff et al.(2002)]{roueff2002} Roueff, E., Felenbok,
P., Black, J.~H., \& Gry, C.\ 2002, \aap, 384, 629

\bibitem[Sato et al.(2010)]{sato2010} Sato, M., Reid, M.~J., Brunthaler,
A., \& Menten, K.~M.\ 2010, \apj, 720, 1055



\bibitem[Scott(1978)]{scott1978} Scott, P.~F.\ 1978, \mnras, 183, 
435

\bibitem[Stantcheva \& Herbst(2003)]{stantcheva03} Stantcheva, T. \&
  Herbst, E. 2003, MNRAS, 340, 983

\bibitem[Teyssier et al.(2004)]{teyssier2004} Teyssier, D., Foss{\'e},
D., Gerin, M., et al.\ 2004, \aap, 417, 135

\bibitem[Ungerechts \& Thaddeus(1987)]{ungerechts1987} Ungerechts, H.,
\& Thaddeus, P.\ 1987, \apjs, 63, 645

\bibitem[van der Tak et al.(2006)]{vandertak2006} van der Tak, F.~F.~S.,
Walmsley, C.~M., Herpin, F., \& Ceccarelli, C.\ 2006, \aap, 447, 1011



\bibitem[Viti et al.(2004)]{viti04} Viti, S., Collings, M.P., Dever,
J.W., McCoustra, M.R.S. \& Williams, David A. 2004, MNRAS, 354, 1141


\bibitem[Whittet et al.(1988)]{whittet1988} Whittet, D.~C.~B., 
Bode, M.~F., Longmore, A.~J., et al.\ 1988, \mnras, 233, 321

\bibitem[Wilson et al.(1970)]{wilson1970}  Wilson, T.~L., Mezger, 
P.~G., Gardner, F.~F. and Milne, D.~K., 1970, \aplett, 5, 99

\bibitem[Wilson \& Rood(1994)]{wilson1994} Wilson, T.~L., \& Rood, R.\
1994, \araa, 32, 191 

\bibitem[Yu et al.(2010)]{yu2010} Yu, S., Pearson,
J.~C., \& Drouin, B.~J.\, et al., 2010, \jcp, 133, 17 

\bibitem[Zhang \& Ho(1997)]{zhang1997} Zhang, Q., \& Ho, P.~T.~P.\ 1997,
\apj, 488, 241

\end{thebibliography}
\end{document}